\documentclass[english,aps,amssymb,prx,twocolumn]{revtex4}
\pdfoutput=1
\usepackage[T1]{fontenc}
\usepackage[latin9]{inputenc}
\usepackage{amsmath}
\usepackage{graphicx}
\usepackage{amssymb}
\usepackage{esint}
\usepackage{tensor}
\usepackage{xfrac}

\makeatletter
\@ifundefined{textcolor}{}
{
 \definecolor{WHITE}{gray}{1}
 \definecolor{RED}{rgb}{1,0,0}
 \definecolor{GREEN}{rgb}{0,1,0}
 \definecolor{BLUE}{rgb}{0,0,1}
 \definecolor{CYAN}{cmyk}{1,0,0,0}
 \definecolor{MAGENTA}{cmyk}{0,1,0,0}
 \definecolor{YELLOW}{cmyk}{0,0,1,0}
 }

\renewcommand{\vec}{\mathbf}
\renewcommand{\vec}[1]{{\bf #1}}

\newcommand{\bs}{\boldsymbol}
\newcommand{\mc}{\mathcal}
\newcommand{\para}{\parallel}

\newcommand{\TG}{\addtocounter{equation}{1}\tag{\theequation}}

\DeclareMathOperator{\sgn}{sgn}

\renewcommand{\phi}{\varphi}
\renewcommand{\epsilon}{\varepsilon}

\usepackage{epsfig}
\makeatother

\usepackage{babel}

\begin{document}

\title { Designer curved-space geometry for relativistic fermions in Weyl metamaterials}
\author{Alex Weststr\"om}
\author{Teemu Ojanen}
\email[Correspondence to ]{teemuo@boojum.hut.fi}
\affiliation{Department of Applied Physics (LTL), Aalto University, P.~O.~Box 15100,
FI-00076 AALTO, Finland }
\date{\today}
\begin{abstract}

Weyl semimetals are recently discovered materials supporting emergent relativistic fermions in the vicinity of band-crossing points known as Weyl nodes. The positions of the nodes and the low-energy spectrum depend sensitively on the time-reversal (TR) and inversion (I) symmetry breaking in the system. We introduce the concept of Weyl metamaterials where the particles experience a 3d curved geometry and gauge fields emerging from smooth spatially varying TR and I breaking fields. The Weyl metamaterials can be fabricated from semimetal or insulator parent states where the geometry can be tuned, for example, through inhomogeneous magnetization. We derive an explicit connection between the effective geometry and the local symmetry-breaking configuration. This result opens the door for a systematic study of 3d designer geometries and gauge fields for relativistic carriers. The Weyl metamaterials provide a route to novel electronic devices as highlighted by a remarkable 3d electron lens effect.  
\end{abstract}

\maketitle
\bigskip{}

\section{Introduction}

Simulating relativistic phenomena in table-top systems has become a major theme in condensed matter physics \cite{volovik}. Topological materials \cite{bernevig}, exhibiting a myriad of connections to high-energy physics \cite{volovik}, have been a significant inspiration for these developments. An important  source of fascination has also been provided by semimetals which display emergent relativistic dynamics at low energies  \cite{katsnelson,hosur,burkov1,wan,lu,xu,liu,liu1,neupane,wang,borisenko,shekhar,xu2,yang,lv,cano}. This has given rise to a wide-spread interest in engineering artificial gauge fields in graphene \cite{katsnelson,vozmediano, amorim} and 3d Dirac  and Weyl semimetals \cite{cortijo1,cortijo2,grushin,pikulin, shapourian, zabolotskiy}. The phenomenology of general relativity and curved-space dynamics has also penetrated into condensed-matter research \cite{volovik}. In addition to fundamental interest, curved-space physics may also have striking practical applications as electromagnetic metamaterials and transformation optics demonstrate \cite{pendry,leonhardt1,leonhardt2}.    

\begin{figure}
\includegraphics[width=0.99\columnwidth]{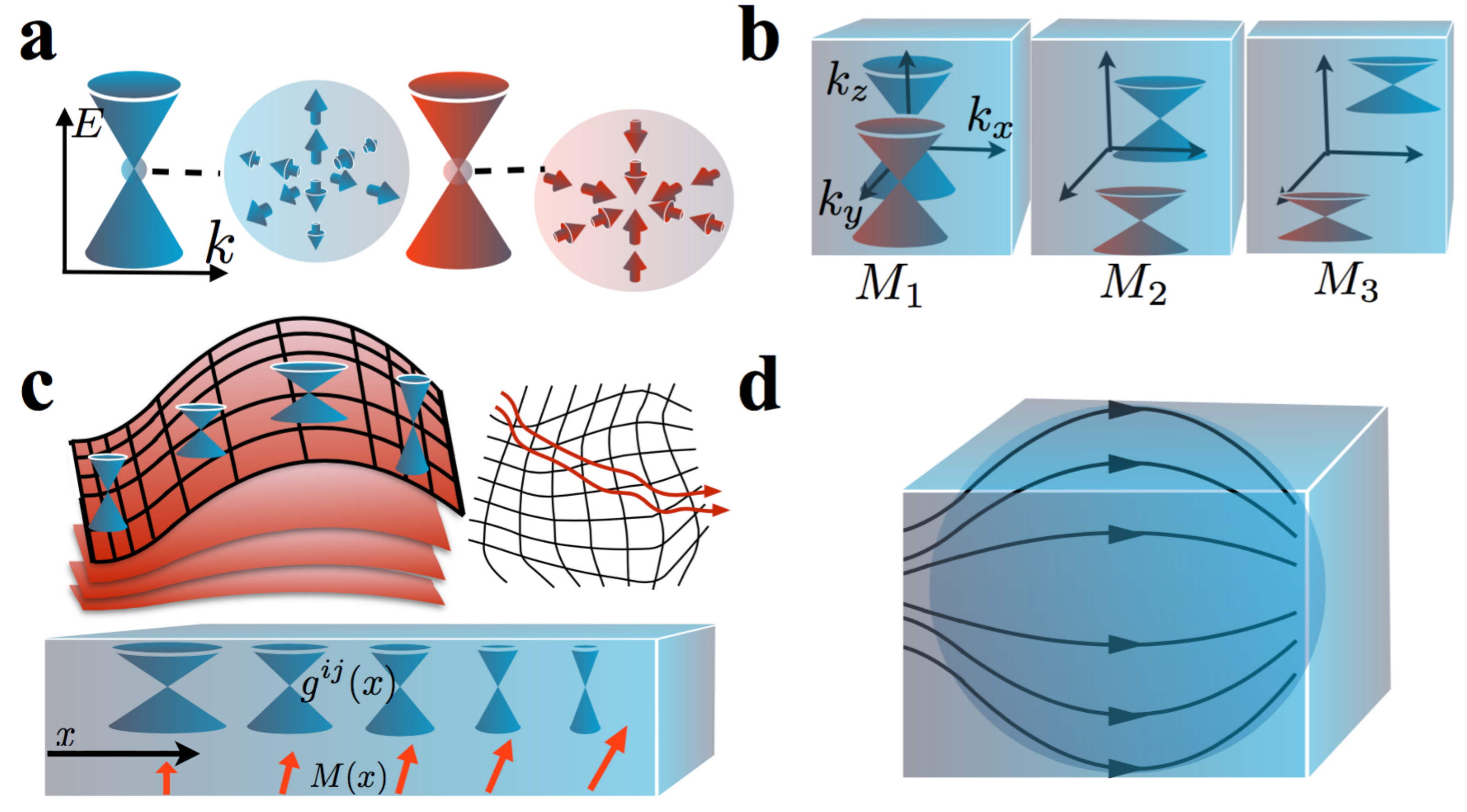} 
\caption{\textbf{a}: Weyl semimetals exhibit one or more band-crossing pairs which act as sources and sinks of the Berry curvature. The low-energy spectrum near the nodes is described by the Weyl Hamiltonian $H=\pm v\vec{k}\cdot \sigma$ resulting in a conical relativistic dispersion.  \textbf{b}: Locations of the Weyl nodes and the the conical envelopes depend on the magnitude of the TR and I breaking in the system and are not fixed by symmetry. For example, by varying the magnetization $M_i$, one can widely tune the spectrum. \textbf{c}: A smoothly varying magnetization in real space (bottom) give rise to local Weyl cones that are deformed depending on the texture. This translates to an effective curved-space geometry experienced by relativistic carriers (top). \textbf{d}: Semiclassical trajectories of carriers in the sample can be engineered by designing suitable TR and I breaking textures.  } \label{fig:schematic}
\end{figure}

In this work, we study a class of structures which we call \emph{Weyl metamaterials}. In these systems, the chiral Weyl fermions familiar from Weyl semimetals, depicted in Fig.~1\textbf{a}, are moving in an artificial 3d curved-space geometry. The great advantage of these systems is that, first of all, the effective geometry experienced by the particles can be widely tuned by external conditions. Since the band crossings in Weyl semimetals result from accidental degeneracy, the positions of the nodes and the low-energy spectrum can be tuned by a large amount by varying time-reversal (TR) and inversion (I) symmetry breaking perturbations, as depicted in Fig.~1\textbf{b}. In addition, a 3d curved geometry can be realized by atomically smooth TR or I breakings whose variation is significant only on the scale of the sample size, as illustrated in Fig.~1\textbf{c}. These two factors combined with the obvious richness offered by 3d space give Weyl metamaterials  flexibility and  enormous advantage over 2d materials \cite{katsnelson,vozmediano, amorim} as a platform for relativistic curved-space physics. Weyl metamaterials can be fabricated from 3d semimetal or insulator parent states by generic inhomogeneous TR and I breaking couplings. Hence the concept reaches substantially beyond the strain-engineering through elastic deformations \cite{cortijo1,cortijo2,grushin,pikulin, shapourian}. More generally, Weyl metamaterials pave the way for novel 3d electronic devices through curvature engineering as sketched in Fig.~1\textbf{d}. In this article, we provide an example of such a device where the Weyl metamaterial acts as a 3d lens for the itinerant charge carriers. A magnetization with a radial gradient in the plane perpendicular to the initial velocities gives rise to an effective curved space and emergent gauge field, which cause the trajectories of incoming charge carriers at different radii to converge at a focal point.

In Sec.~\ref{sec:theory}, starting from a generic four-band model, we derive a curved-space Weyl Hamiltonian $H=\tensor{e}{^\mu_i}(k_\mu-a_\mu)\sigma^i$ describing the low-energy dynamics of Weyl metamaterials. The emergent gauge field $a_\mu$ and the frame fields $\tensor{e}{^\mu_i}$ which encode the metric tensor $g^{\mu\nu}$ are solved as functions of the TR and I breaking fields. These expressions provide the fundamental tool for systematic reverse-engineering of synthetic curved-space geometries and gauge fields in Weyl metamaterials. Our results are applicable to a wide class of different lattice and continuum models as illustrated by examples. The general theory is illustrated in Sec.~\ref{sec:motion} by proposing simple magnetic textures that give rise to remarkable electron lensing effects.
In Sec.~\ref{sec:discussion}, we briefly discuss the experimental feasibility of inhomogeneous symmetry-breaking fields, and in the final section summarize our work and results, as well as propose directions for future research.

\section{Theory of Weyl metamaterials}\label{sec:theory}

\subsection{Engineering Weyl metamaterials from general parent states} 

The central idea of our work is to start from a generic insulating or semimetal phase and introduce smooth spatially varying TR and I breaking (TRIB) fields that push the parent system into an inhomogeneous Weyl semimetal phase. The TR and I breaking terms are not assumed to be small in any sense, however their spatial dependence is assumed to be smooth. Besides an intrinsic Weyl semimetal, the parent state for a Weyl metamaterial could be a Dirac semimetal or TR invariant topological insulator \cite{bernevig,zhang} as it is well known that these systems generically undergo a transition to Weyl semimetals when the I \cite{murakami,ojanen} or TR symmetry is broken. A possible mechanism to break TR in topological insulators is magnetic doping, which has been demonstrated in thin films \cite{chang1,chang2,bestwick}.

As the minimum models of Weyl semimetals have four energy bands, we write down the Clifford representation of the most general TR and I invariant four-band models and identify all TR and I breaking terms as was also done in Ref.~\cite{burkov3}. Without any TRIB fields, the most general four-band Hamiltonian is
\begin{equation}\label{eq:H0}
	H_0 =  n(\vec k)\mathbb{I} + \bs\kappa(\vec k)\cdot\bs\gamma + m(\vec k) \gamma_4,
\end{equation}
where the parameters $n$, and $m$ are symmetric in the momentum $m(\vec k)=m(-\vec k)$, while the kinetic term is antisymmetric $\bs\kappa (\vec k)=-\bs\kappa (-\vec k)$. We have denoted the unit matrix by $\mathbb{I}$. Furthermore, $\gamma_\mu$ denotes the five $4\times 4$ gamma matrices satisfying the anticommutation relations $\{\gamma_\mu,\gamma_\nu\} = 2\delta_{\mu\nu}$. Since the Hamiltonian \eqref{eq:H0} respects TR and I symmetry, it follows that $\gamma_{1,2,3}$ are odd under both I and TR, while $\gamma_4$ must be even. From this, it is then evident that $\gamma_5 = \gamma_1\gamma_2\gamma_3\gamma_4$ must also be odd under both operations. To write down the most general Hermitian $4\times 4$ Hamiltonian, we need to include ten additional matrices constructed using commutators of the gamma matrices, i.e. $\gamma_{ij} \equiv -i[\gamma_i,\gamma_j]/2$. The symmetries of $\gamma_{ij}$ can be easily deduced from the symmetries of their constituent gamma matrices. As it turns out, we can group all ten matrices into four distinct sets: three sets consisting of three mutually anticommuting matrices each, and a fourth set consisting only of the matrix $\gamma_{45}$. The sets, as well as the transformation properties of each set under TR or I, are summarized in table~\ref{tab:transform} -- in fact, we can see that each group breaks either TR or I symmetry. 

The general $4\times 4$ Hamiltonian is expressed as
\begin{equation}\label{eq:fullham}
	H = H_0 + \vec u \cdot \vec b + \vec w \cdot\vec p+ \vec u' \cdot \vec b' + f\epsilon,
\end{equation}
where the functions $\vec u$, $\vec w$, $\vec u'$ and $f$ characterize the TRIB fields and fix the position of the Weyl nodes and the low-energy spectrum. To extract the low-energy Weyl Hamiltonian, we must block diagonalize \eqref{eq:fullham}. To make analytical progress, we concentrate on the case $\vec u'=0$, $f=0$. This restriction leaves room for substantial generality, allowing for a full three-component TR breaking field $\vec u$ and I breaking field $\vec w$. To illustrate the wide applicability of the general formulation, we give in App.~\ref{a:realizations} three concrete examples of Weyl metamaterials: one based on a topological insulator-ferromagnet layer structure \cite{burkov2}, a popular theoretical model describing magnetically-doped topological insulators \cite{vazifeh}, and  a 3d Dirac semimetal \cite{wang,borisenko} with broken inversion symmetry.

 We note that the recently studied photonic Weyl  systems \cite{lu, xiao2} employ a two-mode approximation and, as such, falls outside the scope of the our formalism developed with electronic band structures in mind where the four-band Clifford representation of the Hamiltonian~\eqref{eq:fullham} is the fundamental starting point.

\begin{table}
	\begin{tabular}{|l|r|r|}
		\hline
		 & TR & I\\
		\hline
		$\vec b = (\gamma_{23}\ \gamma_{31}\ \gamma_{12})$ & $-1$ & $+1$ \\
		\hline
		$\vec p = (\gamma_{14}\ \gamma_{24}\ \gamma_{34})$ & $+1$ & $-1$ \\
		\hline
		$\vec b' = (\gamma_{15}\ \gamma_{25}\ \gamma_{35})$ & $-1$ & $+1$ \\
		\hline
		$\varepsilon = \gamma_{45}$ & $+1$ & $-1$ \\
		\hline
\end{tabular}
\caption{The four groups of $\gamma_{\mu\nu}$ and their transformation properties.}\label{tab:transform}
\end{table}

\begin{figure*}
\includegraphics[width=1.99\columnwidth]{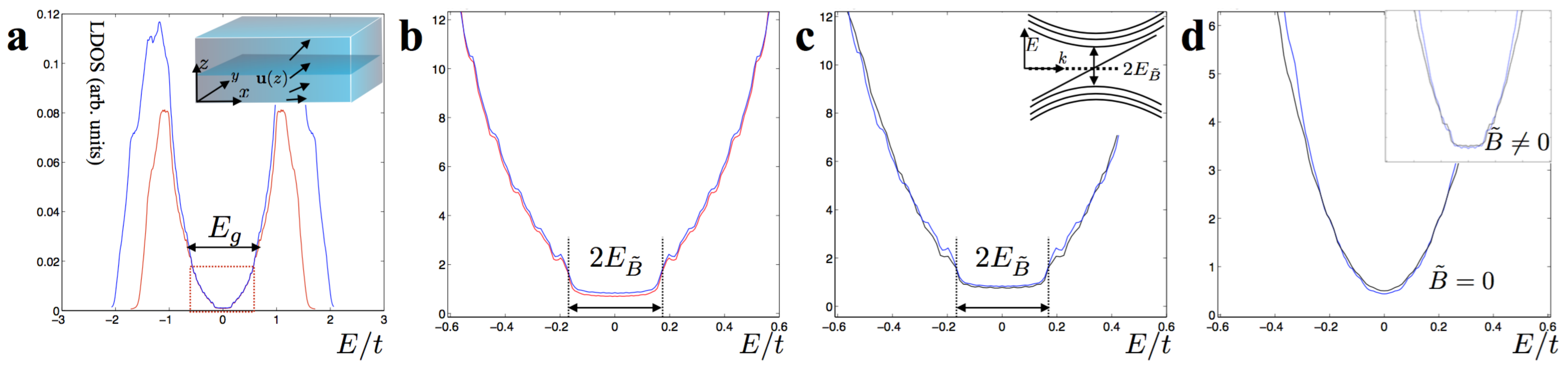} 
\caption{\textbf{a}: LDOS of the local Weyl approximation (red) and the exact four-band Hamiltonian (blue) are in excellent agreement in the energy window $E_g$ around the Weyl node. Results are calculated for the lattice model specified in the text with a rotating TR-breaking texture $\vec{u}(z)=u_0(\cos k_0z,\sin k_0z,0)$. The parameters are $u_0=0.4t$, $k_0=0.05$, $m=0.2t$ and the energy delta functions are smoothed by Lorentzians with width $0.01t$. The inset shows the geometry of the system where the LDOS is calculated. With periodic boundary conditions in the $x$ and $y$ directions, the LDOS depends only on $z$. The curves are plotted for $z=30$ in a system with $N_z=60$ lattice cites.  \textbf{b}: Magnification of the low-energy region marked by the red dashed square in \textbf{a}. The energy scale $E_{\tilde{B}}$ marks the crossover between two trends where LDOS is approximately constant, and where it is parabolic. \textbf{c}: Comparison of the LDOS for the linearized Weyl approximation (black) including both chiralities and the exact four-band Hamiltonian for the same model as in \textbf{a}. Inset: Dispersion of one chiral Weyl fermion in constant magnetic field as a function of parallel wave vector. The constant LDOS region corresponds to the bulk cone gap due to the local effective magnetic field $\tilde{\vec{B}}=\nabla\times \vec{k}^W$. \textbf{d}: LDOS for the linear Weyl approximation (black) and the exact Hamiltonian (blue) for texture $\vec{u}(z)=[0,0,u(z)]$ where $u(z)=u_{1}+(u_{2}-u_1)z/N_z$ with $u_1=0.3t$ and $u_2=0.8t$.  The other parameters are same as in \textbf{a}.  For this system $\tilde{\vec{B}}=0$. Inset: Otherwise the same but  
$\vec{u}(z)=[u(z),0,0]$. Now $ \tilde{\vec B}$ does not vanish and the LDOS is modified around $E=0$ by the effective magnetic field. } \label{fig:LDOS}
\end{figure*}

\subsection{Weyl block reduction}

Here we derive the local Weyl Hamiltonian from the original four-band model \eqref{eq:fullham}. The block reduction is achieved by the unitary transformation theory developed in App.~\ref{a:reduction} and is given by
\begin{equation}\label{eq:blockdiag}
	H' = \begin{pmatrix}
		D_0(\vec k)\sigma^0+\vec D(\vec k)\cdot\bs\sigma & 0\\
		0 & d_0(\vec k)\sigma^0+\vec d(\vec k)\cdot\bs\sigma,
	\end{pmatrix},
\end{equation}
where Weyl nodes are now found within either the upper or lower block, while the other block is gapped. The low-energy physics in the vicinity of the Weyl nodes is contained in the local Weyl approximation $H'_W= d_0(\vec k)\sigma^0+\vec d(\vec k)\cdot\bs\sigma$. Here, we have collected the Pauli matrices $\sigma^{1,2,3}$ into a vector $\bs\sigma$, and written the $2\times 2$ identity matrix as $\sigma^0$. For concreteness, let us relegate the case where $\vec w \neq 0$, $\vec u = 0$, as well as the general case $\vec u \neq 0$, $\vec w \neq 0$ to App.~\ref{a:reduction} and for now consider the specific case $H = \bs\kappa(\vec k)\cdot\bs\gamma + m\gamma_4 + \vec u\cdot\vec b$, which could describe a TR and I invariant insulator  or semimetal in the presence of spatially varying magnetization $\vec M= \vec u$. The local Weyl approximation yields $d_0=0$ and
\begin{equation}\label{eq:d}
	\begin{split}
		d_1& = -\frac{(\hat{\vec u}\times\bs\kappa(\vec k))\cdot\hat{\vec z}}{\sqrt{1-\hat{u}_3^2}}\\
		d_2 &= -\frac{[\hat{\vec u}\times(\bs\kappa(\vec k)\times\hat{\vec u})]\cdot \hat{\vec z}}{\sqrt{1-\hat{u}_3^2}}\\
		d_3 &= u-\sqrt{(\hat{\vec u}\cdot\bs\kappa(\vec k))^2+m^2},
	\end{split}
\end{equation}
where $u=|\vec u|$ and $\hat{\vec u} = \vec u/u$. This system has Weyl nodes $\vec{k}^W$ satisfying $\bs\kappa(\vec k^W) = \pm\sqrt{u^2-m^2}\hat{\vec u}$ whenever $m < |\vec u|$. The gapped block and the results for the general case  $\vec u \neq 0$ and $\vec w\neq 0$ are given in the Appendix. In the local Weyl approximation $H_W'$, the $\vec d$-vector depends not only on $\vec k$ but has an implicit position dependence through $\vec u$ which is slowly varying in space. The Weyl reduction \eqref{eq:blockdiag}, which is exact for position-independent TRIB fields, is an approximation when $\vec u$ and $\vec w$ depend on position. The unitary transformations that block diagonalizes  the four-band model become spatially dependent through $\vec u(\vec{r}), \vec w(\vec{r})$. Regarding $\vec k$ in  $H_0$ as a canonical conjugate operator of $\vec{r}$, the transformation also produces terms proportional to gradients (as well as higher-order spatial derivatives) of  $\vec u(\vec{r}), \vec w(\vec{r})$. The gradient terms could, for example, give rise to off-diagonal elements in Eq.~\eqref{eq:blockdiag}. These terms would give corrections to $H_W'$ of the order of $\mathcal{O}\left[(\partial_i \vec u)^2, (\partial_i \vec w)^2\right]$ which, along with other gradient corrections, are very small for slowly varying fields. This is confirmed by direct numerical comparisons between the local density of states (LDOS) of the exact four-band model and the local Weyl approximation $H_W'$, and typically suggest an excellent agreement between the two. The LDOS, defined by $\nu(\mathbf{r},E)=\sum_{E_n}|\Psi_{n}(\mathbf{r})|^2\delta(E-E_n)$, probes the local properties of the system and therefore is an ideal tool to investigate the local Weyl approximation. A short explanation on how the computation of the LDOS was performed can be found in App.~\ref{a:ldos}. We have calculated the LDOS for representative Weyl-metamaterial models with $H=\bs\kappa(\vec k)\cdot\bs\gamma + m \gamma_4+\vec u \cdot \vec b$, where $\kappa_i(\vec{k})=t\sin{k_i}$ and $m$ and $t$ are constants and different TR-breaking textures $\vec{u}$. This is illustrated in Fig.~2 which establishes an excellent agreement between the exact result and the local Weyl approximation in the energy window $E_g$ determined by the energy gap of the gapped block. Away from the Weyl node but within $E_g$, the smoothed LDOS exhibits a quadratic envelope typical for Weyl semimetals. At the Weyl node, the LDOS typically becomes modified due to the emergent gauge field effect, as clarified in the next section.


\subsection{ Designer curved-space geometry for Weyl particles}

To gain better insight into the low-energy physics, we expand $H_W'$ around the local Weyl points. For each Weyl point, we get a low-energy Hamiltonian of the form
\begin{equation}\label{eq:linear}
	H_{W}' = d_0(\vec k_W)\sigma^0 + \tensor{e}{^j_\nu}(\vec r)\sigma^\nu (k_j-k^W_{j}).
\end{equation}
Here we have employed the Einstein summation convention -- Greek indices represent integers from 0 to 3, while Latin indices take integer values from 1 to 3, unless otherwise stated. We define the frame fields (or tetrads) as
\begin{equation}\label{eq:vierbein}
	\tensor{e}{^i_\nu}(\vec r) = \left.\frac{\partial d_\nu}{\partial k_i}\right\vert_{\vec k = \vec kW},
\end{equation}
where $\vec k^W(\vec{u},\vec{w})$ is the momentum corresponding to the local Weyl point around which we have expanded. Introducing a two-component spinor $\Psi$, the low-energy Schr\"odinger equation then takes the form $H_W'\Psi=\mathcal{E}\Psi$ which when squared can be formally written in the form $g^{\mu\nu}p_\mu p_\nu\Psi = 0$ with appropriately defined canonical momenta $p_\mu$. When the term proportional to the unit matrix vanishes $d_0=0$ in $H_W'$, this equation describes a massless relativistic particle with dispersion $\mathcal{E}^2=g^{ij}p_ip_j$ moving in a curved space.  Here $p_i=k_i-k^W_i$ and the contravariant metric $g^{\mu\nu} = \eta^{ab}\tensor{e}{^\mu_a}\tensor{e}{^\nu_b}$ is defined by the standard relation \cite{zee} using tetrad fields $\tensor{e}{^\mu_\nu}$ which coincide with those defined in Eq.~\eqref{eq:vierbein} with the addition $\tensor{e}{^0_0} = 1$. The covariant metric $g_{\mu\nu}$ is defined as the inverse of $g^{\mu\nu}$ and comes in to play in the semiclassical dynamics through Christoffel symbols.

The frame fields $\tensor{e}{^i_\nu}$ for the model \eqref{eq:d} can be straightforwardly evaluated and are given in App.~\ref{a:reduction}. We can also easily find the inverse (co-frame) elements $\tilde{e}$ that satisfy $\tensor{e}{^\alpha_\nu}\tensor{\tilde{e}}{_\alpha^\mu} = \delta_\nu^\mu$ and $\tensor{e}{^\mu_\alpha}\tensor{\tilde{e}}{_\nu^\alpha} = \delta^\mu_\nu$. The contravariant metric tensor is given by
\begin{equation} 
	\mathfrak g  =
	\begin{pmatrix}
		-1 & 0\\
		0 & \bs g
	\end{pmatrix},
\end{equation}
where the elements of the $3\times 3$ matrix $[\bs g]^{ij}=g^{ij}$ are
\begin{equation} \label{eq:g}
	g^{ij} = (\partial_{k_i}\bs\kappa)\cdot(\partial_{k_j}\bs\kappa)-	\frac{m^2}{u^2}(\partial_{k_i}\bs\kappa\cdot\hat{\vec u})(\partial_{k_j}\bs\kappa\cdot\hat{\vec u}) 
\end{equation}
evaluated at the local Weyl node $\vec k = \vec k^W$. In the form (\ref{eq:g}) we have assumed that $m$ is constant in the vicinity of $\vec{k}^W$, the more general case where $m$ depends on $\vec{k}$ is also given in App.~\ref{a:reduction}. The relations \eqref{eq:d}, \eqref{eq:vierbein}, and \eqref{eq:g}, along with their generalizations given in App.~\ref{a:reduction}, provide the fundamental relation between the effective geometry and the physical symmetry-breaking configuration and serve as a starting point in designing synthetic curved-space geometries for the Weyl quasiparticles. The formula Eq.~\eqref{eq:g} has a similar role in Weyl metamaterials as the relation between the metric and electromagnetic permeability and permitivity tensors in transformation optics \cite{pendry,leonhardt1,leonhardt2}.  It shows concretely how a Weyl metamaterial acts as a transformation media where the geometry is tuned by $\vec{u}$.

We postulate an effective long-wavelength quantum Hamiltonian by making the replacement $\vec k \to -i\nabla$ in Eq.~\eqref{eq:linear}. A direct substitution would leave $H_W'$ non-Hermitian. This can be remedied by the standard prescription of symmetrizing the non-commuting product of gradient and the frame fields. Assuming for simplicity that $d_0=0$, we obtain an effective curved-space Weyl Hamiltonian
\begin{align*}\label{eq:weyl}
	H_W &= \frac{1}{2}\{\tensor{e}{^j_\nu}\sigma^\nu, -i\partial_j-qa_j\} \\
	%
	&= \tensor{e}{^j_\nu}\sigma^\nu\left( -i\partial_j -qa_j-\frac{i}{2}\tensor{\tilde{e}}{_j^\mu}\partial_n\tensor{e}{^n_\mu}\right).\TG
\end{align*}
Here we have defined an effective gauge field $\vec{a}=\vec{k}^W(\vec{r})/q$ -- $q$ being the charge of the quasiparticle -- which gets a small correction due to the symmetrization. Though the correction is parametrically small since it depends on the derivatives of the TRIB fields, we include it to preserve Hermiticity exactly. Symmetrization is a standard way to quantize a product of two noncommuting observables and the minimal procedure to ensure Hermiticity. The situation here is formally analogous to quantizing Hamiltonian with a spatially varying Rashba spin-orbit coupling.  Numerical comparisons show that the effective Hamiltonian \eqref{eq:weyl} produces the low-energy LDOS of the exact four-band model accurately, as seen in Figs. 2\textbf{c} and \textbf{d}. The linear approximation also illuminates the behavior of the exact model. The flat part of the LDOS in the vicinity of the Weyl node appears due to a chiral mode appearing in the gap of the bulk Weyl cones because of the Landau quantization induced by the local effective magnetic field $\tilde{\vec{B}}=\nabla \times \vec {k}^W/q$. The inset of Fig. 2\textbf{c} shows how a constant field of magnitude $|\tilde{\vec{B}}|$ give rise to a chiral mode representing the zeroth Landau level dispersing in the direction the field. The chiral mode has a 1d dispersion and a constant density of states. This interpretation is confirmed by the fact that the calculated LDOS is approximately constant in the energy window $2E_{\tilde{B}}=2v_{\tilde{B}}\sqrt{2q|\tilde{\vec{B}}|}$ which correspond to the gap between $-1$ and $1$ Landau levels  where $v_{\tilde{B}}$ is the Fermi velocity along the magnetic field. In addition, as shown in Fig.~2\textbf{d}, for textures which the effective magnetic field vanishes, the LDOS displays a quadratic trend down to the Weyl node.



\begin{figure*}
\includegraphics[width=1.9\columnwidth]{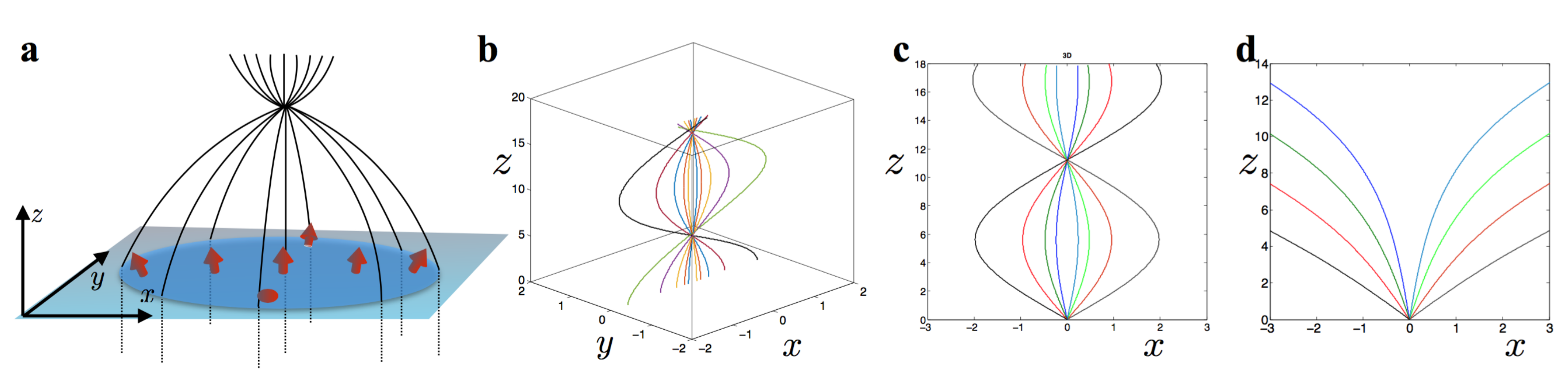} 
\caption{ \textbf{a}: Weyl metamaterial with rotating magnetic texture (red arrows) realizes a system where trajectories of particles with constant energy converge at focal points.     \textbf{b}: Semiclassical trajectories of carriers for the model specified in the main text with parameters $v_x=v_y=v_z=v$, $m/u=0.5$ and $\omega=0.05\pi/\xi$. The numerical values of the coordinates are given in the units of $\xi$ which parametrizes the magnetic rotation. The initial conditions for the trajectories are $y=z=0$, $k_x=k_y=0$, $k_z=1.25u/v$. Different curves correspond to different initial $x$ coordinates. \textbf{c}: Trajectories of the same system but for initial conditions $x=y=z=0$, $k_z=1.25u/v$. Different curves correspond to $k_x=\pm0.02, \pm0.04,\pm0.08,\pm0.16u/v$. \textbf{d}: The same as  \textbf{c} but for the particles of opposite chirality. } \label{fig:two}
\end{figure*}

\section{Particle motion in Weyl metamaterials}\label{sec:motion}

\subsection{Semiclassical dynamics}

To study the quasiparticle dynamics, we employ the semiclassical theory \cite{xiao}. We assume that the system is doped above the Weyl node and that the Landau quantization due to the real and effective gauge fields can be neglected. Then we can ignore interband dynamics and concentrate on a single positive energy band. The semiclassical wave-packet motion for inhomogeneous systems \cite{xiao} is described by the equations of motion  
\begin{align}\label{eq:semiclass}
&\dot{r}^i=\partial_{k_i}\mathcal{E}-\Omega_{k_i r^l}\dot{r}^l-\Omega_{k_i k_l}\dot{k}_l\nonumber\\
&\dot{k_i}=-\partial_{r^i}\mathcal{E}+\Omega_{r^i k_l}\dot{k}_l+\Omega_{ r^i r^l}\dot{ r}^l-q (E_i+\epsilon_{ijk} \dot{r}^jB^k),
\end{align}
where  $\mathcal{E}=(g^{ij}k_ik_j)^{\frac{1}{2}}$ is the relativistic curved-space dispersion and $E$ and $B$ are electric and magnetic fields including the emergent field $\tilde{B}$. Quantum particles have orbital degrees of freedom encoded in the frame fields, leading to quantum-geometric effects given by the Berry forces on top of the classical effects. The novelty compared to the previously extensively studied flat case is that, in addition to the curved-space dispersion, we must consider a $6\times6$ phase-space Berry-curvature tensor $\Omega$ instead of its $3\times3$ momentum block. In the studied model, the Berry tensor has components $\Omega_{q_1 q_2}=-\frac{\epsilon^{ijk}}{2|d|^3}d_i\partial_{q_1}d_j\partial_{q_2}d_k$, where $d_i$ are the coefficients of the Pauli matrices in Eq.~\eqref{eq:blockdiag}. In the linear approximation, the Berry curvatures can be expressed in terms of the frame fields and their derivatives by substituting $d_j=\tensor{e}{^i_j}k_i$. In the lowest non-trivial order, we can drop the $\Omega_{ r^i r^l}$ term since it is the only term that involves two spatial derivatives and is parametrically small. Away from the Weyl nodes, the Berry curvatures vanish as $\Omega_{k_i k_l}\sim \frac{1}{|k|^2}$, $\Omega_{r^i k_l}\sim \frac{1}{|k|}$, so by increasing the doping, one recovers the classical geometric effects together with the emergent magnetic field. However, in general, both classical and quantum geometric effects need to be considered together. On the other hand, we have shown in App.~\ref{a:semiclassical} that in the absence of external electromagnetic fields $E=0,\ B=0$, the ballistic motion is governed by the geodesic equation for a charged particle $\ddot{r}^l + \tensor{\Gamma}{^l_i_j}\dot{r}^i\dot{r}^j = qg^{li}\varepsilon_{ijk}\dot{r}^j\tilde B^k/\mc E$ in the lowest nontrivial order in spatial derivatives. Here $ \tensor{\Gamma}{^l_i_j}$ is the Christoffel symbol calculated from the metric $g_{ij}$. The geodesic motion is a direct manifestation of the curved geometry and indicates how the carrier motion can be tailored through geometry. The carrier dynamics in disordered solids also include scattering processes, which could be implemented by incorporating the semiclassical dynamics into a Boltzmann equation. However, this goes beyond the scope of the present work where we illustrate the designer geometries by solving  Eq.~\eqref{eq:semiclass}.

\subsection{ 3D Weyl electron lens} 

The negative index of refraction in optical metamaterials enable fabrication of  exotic devices \cite{leonhardt2} such as the Veselago lens \cite{veselago} which can also be realized for electrons in graphene \cite{cheianov}. Inspired by these ideas, we propose a 3d Weyl electron lens as an application of Weyl metamaterials. This phenomenon can be realized, for example, in a two-node model $H=v_ik_i\gamma_i+ m \gamma_4+\vec u \cdot \vec b$, which could represent a topological insulator-ferromagnet sandwich structure or magnetically-doped topological insulator with slowly rotating magnetization $\vec{u}=u\left[\cos\phi\sin\theta(r) ,\sin\phi\sin\theta(r), \cos \theta(r)\right]$ expressed in spherical coordinates. The texture is parametrized by the function $\theta(r)=\omega r$ and depicted in Fig.~3 \textbf{a} which illustrates the carrier trajectories in the setup. The carriers of one chirality entering the sample in the vicinity of $r=0$ will converge periodically in the vicinity of focal points. This motion results from a combined effect of the curved geometry and the effective magnetic field that can be straightforwardly calculated in our general framework. In Fig.~3 \textbf{b} we have solved the trajectories from the semiclassical equations corresponding to initial conditions where parallel beams enter the sample at $z=0$. Thus, the image of the object at infinity is reproduced at focal points. The situation in Fig.~3 \textbf{c} corresponds to the case where a point-like object is located at $z=0$ and imagined at the focal points. The carriers with the opposite chirality experience an opposite magnetic field and diverge away from $x=y=0$ axis as shown in Fig.~3 \textbf{c}. Thus the considered geometry will realize a chirality-selective 3d electron lens. This effect is similar to the chirality-dependent Hall effect discovered recently in inhomogeneous Weyl semimetals \cite{yang2}.  

It can be shown analytically (see App.~\ref{a:analytical}) that for small-amplitude oscillations, assuming $\sqrt{k_x^2 + k_y^2}/k_z \ll 1$, $v_x=v_y=v$, the distance between focal points $\Delta z$ is
\begin{equation}
	\Delta z = 2\pi\frac{v_z}{v\omega}\sqrt{\frac{1-\frac{m^2}{u^2}}{\frac{m^2}{u^2}+\frac{u}{v_zk_z}\sqrt{1-\frac{m^2}{u^2}}}},
\end{equation}
if we neglect the effects of the Berry curvature -- the exact numerical solution shows that the motion is not completely planar due to quantum geometric effects. In the absence of the effective magnetic field, the deviation $\delta l$ from the plane when the particle traverses distance $\Delta z$ along the $z$ axis can be shown from the Berry curvatures together with Eqs.~\eqref{eq:semiclass} to scale as $\delta l/\Delta z \sim \omega k_\para/k_z^2$, where $k_\para$ stands for the momentum within the plane, perpendicular to the $z$ axis. In the presence of the effective magnetic field, this scaling relation is less exact in its momentum dependence, but nevertheless provides the correct order of magnitude for the deviation. Thus, in the regime of interest it holds that $\delta l\ll \Delta z$ so the lens motion is nearly planar. The expression for $\Delta z$ shows how the length scale for the lensing effect is controlled by the rotation of the texture through $1/\omega$. An arbitrarily slowly varying texture would yield a similar lens effect with a longer separation of the focal points. The lens effect depends on the energy of the incoming particles only weakly through $k$ in the denominator of the square root.

Interestingly, a similar 3d lens effect can also be realized in unidirectional magnetic textures as discussed in App.~\ref{a:lenses}. We also discovered a 2d lens effect for a simpler magnetic texture that only rotates in plane.  As smoking-gun observable evidence of the lens phenomenon one could consider the quantum mirage effect pointed out in Ref.~\cite{cheianov}: a point-like impurity will in general give rise to localized Friedel oscillations whose image due to the lens geometry can be observed away from the impurity, giving rise to a mirage perturbation. For example, a density variation due to an impurity located at the origin in Fig.~3 c would have its image appear at the convergence of the trajectories near $(0,0,11)$. The 3d Weyl lens illustrates that Weyl metamaterials posses richness beyond the 2d materials considered for curvature and gauge field engineering. Interesting future challenges include, for example, finding realizations for 3d electronic invisibility devices.

In both the numerical and analytical analyses of the 3D lens geometry, we have not considered surface effects that might occur when injecting wave packets through sharply defined boundaries. To study sharp interface effects, one should complement the present theory of smooth TRIB fields with additional boundary conditions. This will be addressed in future work. Nevertheless, the TRIB-field configuration will result in oscillatory behavior for one chirality of the incident particles once they are inside the metamaterial.
 
\section{Discussion}\label{sec:discussion}

Experimental realizations of the proposed Weyl metamaterial structures require controlled fabrication of systems with inhomogeneous TR or I breaking.  Weyl-metamaterial structures can be fabricated from topological insulators and topological semimetals -- both of which have an ever-growing number of experimental realizations. In App.~B, we have outlined how four much-studied models where uniform TR and I breaking has been promoted to inhomogeneous can serve as platforms for Weyl metamaterials. Local breaking of inversion through strain is already considered feasible \cite{grushin, pikulin} and, considering the relentless experimental advances in the field, local magnetic manipulation will become accessible in the near future. For example, artificially constructed layered Weyl semimetals \cite{burkov2} are now becoming experimentally realizable \cite{hasan}; artificially grown layer structures allow a degree of manipulation of the magnetization in the magnetic layers. Another interesting option for realizing varying magnetization arises from the natural magnetization dynamics in magnetically doped topological semimetals. These systems are predicted to display textured ground states or excitations \cite{araki} which could be employed in metamaterial structures. 

Now we turn to the question of experimental requirements important in realizing functional Weyl metamaterial structures. Considering the relatively mild requirements for physical parameters and the rapid advance of experimental methods, we are optimistic that the first Weyl metamaterial structures can be fabricated in the near future. It is promising that the required rate of variation of the symmetry breaking fields do not need to exceed specific critical value. The characteristic scale of variation simply fixes the overall size of functional units. For example, reducing the rate of rotation of the magnetization in the 3d lens geometry increases the focal length proportionally.  To be more precise and to give simple estimates to guide the experimental realizations, it is illuminating to consider a two-node example such as the Weyl lens Hamiltonian.  In order to observe the effects of the curvature and the emergent gauge field, their length scales must be comparable or smaller than the length scale of the sample, i.e. for a sample with linear dimensions $L$, we have $|\partial_i \vec{u}|L/u\gtrsim 1$, where $u$ and $|\partial_i \vec{u}|$  are the characteristic magnitudes of the symmetry-breaking field $\vec{u}$ and its gradient in the sample volume. On the other hand, the Weyl description breaks down if the spatial variation of the magnetic texture is too large, enabling scattering between different Weyl nodes. For Weyl nodes separated by a distance $\Delta k^W$, the spatial variation must hence satisfy $|\partial _i\vec{u}|/(\Delta k^Wu) \ll 1$.  The two requirements can in general be satisfied in systems where $\Delta k^W \gg  L^{-1}$, which is not a serious restriction for Weyl metamaterial structures. In the particular case of the Weyl lens, we can estimate that $|\partial_i \vec{u}|/u=\omega$, leading to constrains $\omega L\gtrsim1$ and $\omega\ll v_F\sqrt{u^2-m^2}$. We see that the requirements can be satisfied in a sufficiently large sample as long as the system is in the Weyl semimetal phase ($u>m$). Of course, in realistic systems there could be other scales affecting the performance such as the momentum relaxation length that need to be taken into account. Nevertheless, there are no fundamental restrictions for the size of the gradients of the magnetic textures and inversion breaking other than being slow enough as to not allow for scattering between Weyl nodes. Therefore it seems  plausible that simple metamaterial structures could be realized in the near future.

Furthermore, the discussion in the previous paragraph points to an important feature of Weyl semimetals not only restricted to the lens geometry. Namely, the curved-space effects in Weyl metamaterials follow almost exclusively from the gradients of the TRIB fields, while the role of the absolute strengths of the fields is only minor in this regard. Certainly, in order for the Weyl-metamaterial picture to be valid, the local field strengths must be such that a corresponding homogeneous sample would be in a Weyl semimetal phase. So any restrictions put on the field strengths in a Weyl metamaterial are the same restrictions as for any corresponding Weyl semimetal in the same sample. Put slightly differently: any physical realization of a Weyl semimetal in an electronic system should in principle also be tunable into a Weyl metamaterial.

\section{conclusion}\label{sec:conclusion}

In this work, we proposed Weyl metamaterials as a highly tunable platform for relativistic fermions in curved space. Analogous to optical metamaterials, we show that Weyl metamaterials offer a 3D electronic platform where local manipulations of TR and I breaking fields allow efficient control over the particle propagation. We laid the ground work for the theoretical description of these systems and established the explicit connection between the local TR and I breaking, and the effective geometry and gauge fields experienced by the carriers. 

From the point of view of applications, Weyl metamaterials offer a novel route to 3D electronic devices. The functionality of the devices arise from the curved-space geometry and emergent gauge fields and they can be systematically designed with the help of our theory framework. In the present work, we provided a concrete example, namely the 3D Weyl electron lens. This structure, realized by relatively simple magnetic textures, focuses carriers depending on their chirality. This is just the first example of the new possibilities offered by Weyl metamaterials. An interesting venue for future work is to study what other exotic phenomena can be engineered in these systems. For example, a possibility to realize electronic cloaking devices from Weyl metamaterials is particularly intriguing.    

From the fundamental point of view, our results can be employed in systematic reverse-engineering of 3D curved geometries and studying quantum effects \cite{birrell} in designer geometries. One direction is then to study the curved-space modifications to the exotic response properties such as the chiral anomaly and its manifestations in Weyl semimetals \cite{liu2,huang,xiong,li,zhang2,behrends}.  Also, fabrication of condensed-matter analogues of cosmological horizons could bring new fundamental phenomena within the reach of experimental studies. Considering the rapid developments in materials and realizations of topological semimetals, the different aspects of curved-space quantum physics will likely become experimentally accessible soon.

\acknowledgements The authors acknowledge the Academy of Finland for support.

\appendix

\section{Weyl block reduction}\label{a:reduction}

\subsection{TR breaking without I breaking}
In this supplement, we derive the unitary transformations that reduce the $4\times 4$ Hamiltonians for different models into blocks of ($2\times 2$) Weyl Hamiltonians. 

It turns out that it is beneficial to first consider the Hamiltonian in Eq.~(2) of the main text with $\vec w = 0$, i.e.
\begin{equation}\label{app:eq:hu}
	H = \bs\kappa(\vec k)\cdot\bs\gamma + m(\vec k)\gamma_4 + \vec u\cdot\vec b.
\end{equation}
As can be easily shown by inspection, the matrices in $\vec b$ have identical properties to those of the regular $2\times 2$ Pauli matrices, i.e.
\begin{equation}
	\{b_i,b_j\} = 2\delta_{ij},\quad [b_i,b_j] = 2i\varepsilon_{ijk}b_k.
\end{equation}
These matrices, like the Pauli matrices, thus exponentiates to the special unitary group SU(2), which is two-to-one homomorphic to the orthogonal group O(3). In practice, this means that the transformation
\begin{equation}
	\vec u\cdot \vec b\to e^{i\frac{\phi}{2}\hat{\vec n}\cdot\vec b}(\vec u\cdot\vec b)e^{-i\frac{\phi}{2}\hat{\vec n}\cdot\vec b} 
\end{equation}
corresponds to rotating the vector $\vec u$ by an angle $\phi$ around the axis defined by the unit vector $\hat{\vec n}$. Furthermore, by observing that the components of the other vectors in table~1 in the main text along with $\bs \gamma$ can be written as $\vec p = -\gamma_5\vec b$, $\vec b' = \gamma_4\vec b$, and $\bs\gamma = -\gamma_{45}\vec b$,  respectively, we can conclude that the exponential map has the same effect (it commutes with $\gamma_4$ and $\gamma_5$) on terms containing these vectors as well.

The first step is to rotate $\vec u$ in Eq.~\eqref{app:eq:hu} to point along the $x$ axis. This can for example be achieved by first rotating $\vec u$ into the $xz$ plane and then rotating it around the $y$ axis, aligning it with the $x$ axis. These two transformations are effected by
\begin{equation}
	\mc U(\vec u) = e^{i b_2\frac{(\theta-\pi/2)}{2}}e^{i b_3 \frac{\varphi}{2}},
\end{equation}
which is expressed in terms of the polar ($\theta$) and azimuthal ($\varphi$) angles of $\vec u$. Applying this transformation to the whole Hamiltonian yields
\begin{equation}\label{app:eq:UHU}
	\begin{split}
		\mc U(\vec u)H\mc U^\dag(\vec u) = (\bs\kappa(\vec k)\cdot\hat{\vec u})\gamma_1 + \frac{(\hat{\vec u}\times\bs\kappa(\vec k))\cdot\hat{\vec z}}{\sqrt{1-\hat{u}_3^2}}\gamma_2\\
		+ \frac{[\hat{\vec u}\times(\bs\kappa(\vec k)\times\hat{\vec u})]\cdot \hat{\vec z}}{\sqrt{1-\hat{u}_3^2}}\gamma_3 + m(\vec k)\gamma_4 + u\gamma_{23}.
	\end{split}
\end{equation}
To block-diagonalize this, we need to eliminate either $\gamma_1$ or $\gamma_4$. This is achieved using $\mc R = e^{-i\frac{\rho}{2}\gamma_{14}}$ with $\tan\rho = (\bs\kappa(\vec k)\cdot\hat{\vec u})/m$. The final result is then
\begin{equation}\label{app:eq:RUHUR}
	\begin{split}
		&\mc R\mc U(\vec u)H\mc U^\dag(\vec u)\mc R^\dag =  u\gamma_{23} + \frac{(\hat{\vec u}\times\bs\kappa(\vec k))\cdot\hat{\vec z}}{\sqrt{1-\hat{u}_3^2}}\gamma_2\\
		 &+ \frac{[\hat{\vec u}\times(\bs\kappa(\vec k)\times\hat{\vec u})]\cdot \hat{\vec z}}{\sqrt{1-\hat{u}_3^2}}\gamma_3 + m(\vec k)\sqrt{1+\frac{(\bs\kappa(\vec k)\cdot\hat{\vec u})^2}{m^2(\vec k)}}\gamma_4,
	\end{split}
\end{equation}
which splits into blocks in the basis given by for example
\begin{equation}\label{app:eq:gammarep}
	\begin{split}
		\gamma_1 &= \tau_x,\qquad \gamma_2 = \tau_z\sigma_x\\
		\gamma_3 &= \tau_z\sigma_y,\quad \gamma_4 = \tau_z\sigma_z.
	\end{split}
\end{equation}

With this choice, the $\vec d$ vectors for the upper and lower block become
\begin{equation}
	\begin{split}
		d^\pm_1 &= \pm \frac{(\hat{\vec u}\times\bs\kappa(\vec k))\cdot\hat{\vec z}}{\sqrt{1-\hat{u}_3^2}}\\
		d^\pm_2 &=  \pm\frac{[\hat{\vec u}\times(\bs\kappa(\vec k)\times\hat{\vec u})]\cdot \hat{\vec z}}{\sqrt{1-\hat{u}_3^2}}\\
		d^\pm_3 &=  u \pm m(\vec k)\sqrt{1+\frac{(\bs\kappa(\vec k)\cdot\hat{\vec u})^2}{m^2(\vec k)}}
	\end{split}
\end{equation}

Depending on the sign of $m(\vec k)$, the nodes will either be located in the upper or the lower block. The $\vec d$ vector corresponding to the Weyl block will regardless have the form
\begin{equation}
	\begin{split}
		d^W_1 &= \frac{(\hat{\vec u}\times\bs\kappa(\vec k))\cdot\hat{\vec z}}{\sqrt{1-\hat{u}_3^2}}\\
		d^W_2 &= \frac{[\hat{\vec u}\times(\bs\kappa(\vec k)\times\hat{\vec u})]\cdot \hat{\vec z}}{\sqrt{1-\hat{u}_3^2}}\\
		d^W_3 &=  u - \sqrt{m(\vec k)^2+(\bs\kappa(\vec k)\cdot\hat{\vec u})^2}
	\end{split},
\end{equation}
where the potential minus sign on the first and second component have been taken care of using a unitary transformation $\sigma_z$. The above vector vanishes whenever all three components are zero, i.e.
\begin{equation}
	\bs\kappa(\vec k^W) = \sqrt{u^2-m^2(\vec k^W)^2}\hat{\vec u}.
\end{equation}
Furthermore, note that if $\vec d^W(\vec k^W) = 0$, it follows immediately that $\vec d^W(-\vec k^W) = 0$, which is just a manifestation of the fact that Weyl nodes come in pairs.

To arrive at an effective curved-space Weyl Hamiltonian, we linearize $\vec d^W$ around the Weyl nodes. The Weyl Hamiltonian can then be written in terms of frame fields as
\begin{equation}
	H^W = \tensor{e}{^j_i}(k_j-k_j^W)\sigma^i,
\end{equation}
where the frame fields $\tensor{e}{^j_i}$ can be found to be
\begin{align}
	\tensor{e}{^j_1} &= \left.\frac{(\hat{\vec u}\times\partial_{q_j}\bs\kappa(\vec q))\cdot\hat{\vec z}}{\sqrt{1-\hat{u}^2_3}}\right|_{\vec q = \vec k^W}\\
	\tensor{e}{^j_2} &= \left.\frac{\left[\hat{\vec u}\times(\partial_{q_j}\bs\kappa(\vec q)\times\hat{\vec u})\right]\cdot\hat{\vec z}}{\sqrt{1-\hat{u}^2_3}}\right|_{\vec q = \vec k^W}\\
	\tensor{e}{^j_3} &= -\left.\frac{m(\vec{k^W})}{u}\partial_{q_j}m(\vec q)\right|_{\vec q = \vec k^W}\\
	&\mp \left.\sqrt{1-\frac{m^2(\vec k^W)}{u^2}}\partial_{q_j}(\bs\kappa(\vec q)\cdot\hat{\vec u})\right|_{\vec q = \vec k^W},
\end{align}
where the sign in the last component corresponds to either $\vec k^W$(upper sign) or $-\vec k^W$(lower sign). The co-frame fields $\tilde e$ are easy to evaluate whenever $\tensor{e}{^j_0} = \tensor{e}{^0_j} = 0$, and are given by
\begin{equation}
	\tensor{\tilde e}{_l^i} = \frac{1}{2}\frac{\varepsilon^{ijk}\varepsilon_{lmn}\tensor{e}{^m_j}\tensor{e}{^n_k}}{\varepsilon_{stu}\tensor{e}{^s_1}\tensor{e}{^t_2}\tensor{e}{^u_3}},\quad \tensor{\tilde e}{_0^0} = 1.
\end{equation}

Finally, the metric tensor can be straightforwardly calculated to be
\begin{widetext}
\begin{equation}
		g^{ij} = (\partial_{k_i}\bs\kappa)\cdot(\partial_{k_j}\bs\kappa)-	\frac{m^2}{u^2}(\partial_{k_i}\bs\kappa\cdot\hat{\vec u})(\partial_{k_j}\bs\kappa\cdot\hat{\vec u}) \pm \frac{m}{u}\sqrt{1-\frac{m^2}{u^2}}\left((\partial_{k_i}m)(\partial_{k_j}\bs\kappa\cdot\hat{\vec{u}})+(\partial_{k_j}m)(\partial_{k_i}\bs\kappa\cdot\hat{\vec{u}})\right) + \frac{m^2}{u^2}(\partial_{k_i}m)(\partial_{k_j}m),
\end{equation}
\end{widetext}
where we have have omitted writing out the arguments. The whole expression must then be evaluated at the relevant Weyl node.

\subsection{I breaking with preserved TR}

In this section we show how to obtain similar results for a system with only inversion-symmetry breaking. The starting Hamiltonian is
\begin{equation}\label{app:eq:hp}
	H = \bs\kappa(\vec k)\cdot\bs\gamma + m(\vec k)\gamma_4 + \vec w\cdot\vec p.
\end{equation}
The last term can be rotated using $\mc U(\vec w)$ to give
\begin{equation}
	\mc U(\vec w)H\mc U^\dag(\vec w) = \vec K\cdot\bs\gamma + m(\vec k)\gamma_4 + w\gamma_{14},
\end{equation}
where $\vec K$ can be directly extracted from Eq.~\eqref{app:eq:UHU}. On this we can then simply apply a rotation $\exp(i\nu\gamma_{23}/2)$ with $\tan\nu = K_3/K_2$ to get
\begin{equation}
	\begin{split}
		e^{i\frac{\nu}{2}\gamma_{23}}\mc U(\vec w)H\mc U^\dag(\vec w)e^{-i\frac{\nu}{2}\gamma_{23}} &= K_1\gamma_1 + K_2\sqrt{1+\frac{K_3^2}{K_2^2}}\gamma_2\\
		&+ m(\vec k)\gamma_4 + w\gamma_{14}.
	\end{split}
\end{equation}
Employing, for example, the representation
\begin{equation}
	\begin{split}
		\gamma_1 = \tau_z\sigma_x,\quad \gamma_2 = \tau_z\sigma_z,\quad \gamma_3 = \tau_x,\quad \gamma_4 = \tau_z\sigma_y,
	\end{split}
\end{equation}
we see that this Hamiltonian is block diagonal with blocks given by
\begin{equation}
	H_\pm = \pm K_1\sigma_x + \left(w \pm K_2\sqrt{1+\frac{K_3^2}{K_2^2}}\right)\sigma_z \pm m(\vec k)\sigma_y.
\end{equation}

It follows that the Weyl block is 
\begin{equation}
	H_W = (\hat{\vec w}\cdot\bs\kappa(\vec k))\sigma_x + \left(w -\sqrt{ K_2^2+K_3^2}\right)\sigma_z + m(\vec k)\sigma_y,
\end{equation}
where, if $\sgn(K_2(\vec k^W)) < 0$, we apply a transformation $\sigma_z$ to get rid of the minuses in front of $K_1 = \hat{\vec w}\cdot\bs\kappa(\vec k)$ and $m$. 
The condition for the existence of the Weyl nodes $\vec{k}^W$ become 
\begin{align}
&m(\vec{k}^W)=0,\nonumber\\
&\bs{\kappa}(\vec{k}^W)\cdot \vec{w}=0,\nonumber\\
&|\bs{\kappa}(\vec{k}^W)|= |\vec{w}|.
\end{align}
In this case, the minimum number of nodes is four.  We can now linearize this around the Weyl points, yielding
\begin{equation}
	\begin{split}
		H_W \approx {\bigg[}&\hat{\vec w}\cdot\partial_{q_j}\bs\kappa(\vec q)\vert_{\vec q = \vec k^W}\sigma_x + \partial_{q_j}m(\vec q)\vert_{\vec q = \vec k^W}\sigma_y\\
		&- w^{-1}\left.\bs\kappa(\vec k^W)\cdot\partial_{q_j}\bs\kappa(\vec q)\right\vert_{\vec q = \vec k^W}\sigma_z {\bigg]}(k_j - k^W_j).
	\end{split}
\end{equation}
We can now immediately read off the frame fields to be
\begin{align}
	\tensor{e}{^j_1} &= \hat{\vec w}\cdot\partial_{q_j}\bs\kappa(\vec q)\vert_{\vec q = \vec k^W}\\
	\tensor{e}{^j_2} &= \partial_{q_j}m(\vec q)\vert_{\vec q = \vec k^W}\\
	\tensor{e}{^j_3} &= \left.\frac{\bs\kappa(\vec k^W)\cdot\partial_{q_j}\bs\kappa(\vec q)}{w}\right\vert_{\vec q = \vec k^W},
\end{align}
and the metric tensor
\begin{equation}
	\begin{split}
		g^{ij} = &(\hat{\vec w}\cdot\partial_{q_i}\bs\kappa(\vec q))(\hat{\vec w}\cdot\partial_{q_j}\bs\kappa(\vec q)) + \partial_{q_i}m(\vec q)\partial_{q_j}m(\vec q)\\
		&+ \frac{\bs\kappa(\vec k^W)\cdot\partial_{q_i}\bs\kappa(\vec q)}{w}\frac{\bs\kappa(\vec k^W)\cdot\partial_{q_j}\bs\kappa(\vec q)}{w},
	\end{split}
\end{equation}
where all terms should be evaluated at $\vec q = \vec k^W$.

\subsection{Block reduction with TR and I breaking}

The problem becomes significantly more cumbersome when both $\vec u$ and $\vec w$ are non-vanishing. In this case, it is useful to write $\vec w = \vec w_\perp + \vec w_\para$, where $\vec w_\perp\cdot\vec u = 0$ and $\vec w_\para\para\vec u$. The goal is to find a transformation that removes all matrices belonging to $\vec p$ from the Hamiltonian, such that we are left with the special case discussed in the first part of this appendix.

Before tackling the full problem, first consider the case where $\vec w_\perp = 0$. Then it follows that $\mc U(\vec u)$ also aligns $\vec w$ with the $x$ axis. The Hamiltonian is then
\begin{equation}
	\mc U(\vec u)H\mc U^\dag(\vec u) = \vec K\cdot\bs\gamma + m\gamma_4+u\gamma_{23} + w\gamma_{14},
\end{equation}
where $\vec K$ can be directly extracted from Eq.~\eqref{app:eq:UHU}. We can now remove $\gamma_{14}$ by a simple unitary transformation $\mc F = e^{i\delta\gamma_1/2}$, provided that $\tan\delta = w/m$:
\begin{equation}
	\begin{split}
		\mc F\mc U(\vec u)H\mc U^\dag(\vec u)\mc F^\dag = K_1\gamma_1 + \cos\delta(K_2\gamma_2+K_3\gamma_3)\\
		+ m\sqrt{1+\frac{w^2}{m^2}}\gamma_4 + ub_1 + \sin\delta(b_2K_3-b_3K_2).
	\end{split}
\end{equation}
Evidently, $\mc F$ also introduces new $\vec b$ terms, but at this point, we can simply refer back to the first half of this appendix, as the problem is mathematically identical to the case $\vec w = 0$. 

Solving the full problem can thus be done if we can find a transformation that aligns the vector associated with $\vec b$ and the vector associated with $\vec p$. For the sake of clarity, we will not concern ourselves with that of $H_0$ until at the very end (as it turns out, all transformations merely rotates $\bs\kappa$), so our primary focus will now lie on
\begin{equation}
	H_1 = \vec u\cdot\vec b + \vec w_\para\cdot\vec p + \vec w_\perp\cdot\vec p.
\end{equation}
As usual, we begin by applying $\mc U(\vec u)$, which gives us
\begin{align*}
	\mc U H_1 \mc U^\dag = ub_1 + \sgn(\vec w\cdot\vec u)w_\para p_1 + \frac{(\hat{\vec u}\times\vec w_\perp)\cdot\hat{\vec z}}{\sqrt{1-\hat{u}_3^2}}p_2\\
	+ \frac{\vec w_\perp\cdot\hat{\vec z}}{\sqrt{1-\hat{u}_3^2}}p_3 \equiv ub_1 + w_\para p_1 + \tilde w_{\perp,2}p_2 + \tilde w_{\perp, 3}p_3\TG.
\end{align*}
We can then make a rotation around $b_1$ with the unitary transformation $\mc V = e^{i\mu b_1/2}$ which for $\tan\mu = \tilde w_{\perp, 3}/\tilde w_{\perp, 2}$ eliminates the $p_3$ term:
\begin{equation}
	\mc V\mc U H_1 \mc U^\dag\mc V^\dag = ub_1 + \sgn(\vec w\cdot\vec u)w_\para p_1 + w_\perp p_2.
\end{equation}
At this point, we make two transformations in succession; we first apply $\mc V_b = e^{i\alpha b_3/2}$ and then $\mc V_p = e^{i\beta p_3/2}$ which gives a seemingly unwieldy
\begin{widetext}
\begin{align*}
	\mc V_p\mc V_b\mc V\mc U H_1 \mc U^\dag\mc V^\dag\mc V_b^\dag\mc V_p^\dag = &\left[u\cos\beta\cos\alpha + \sin\beta(w_\perp\cos\alpha - w_\para'\sin\alpha)\right]b_1 + \left[\cos\beta(w_\para'\cos\alpha+w_\perp\sin\alpha)-u\sin\alpha\sin\beta\right]p_1\\
	- &\left[u\cos\beta\sin\alpha +\sin\beta(w_\para'\cos\alpha+w_\perp\sin\alpha)\right]b_2 - \left[u\sin\beta\cos\alpha -\cos\beta(w_\perp\cos\alpha-w_\para'\sin\alpha)\right]p_2,
\end{align*}
\end{widetext}
where $w_\para' = \sgn(\vec w\cdot\vec u)w_\para$. We can now determine the angles $\alpha$ and $\beta$ by requiring that the terms proportional to $b_2$ and $p_2$ vanish. One possible solution is then
\begin{equation}
	\begin{split}
		\tan\alpha &= \frac{u^2+w_\perp^2 - w_\para^2}{2w_\perp w_\para} - \sqrt{\frac{(u^2+w_\perp^2 - w_\para^2)^2}{4w_\perp^2 w_\para^2} + 1}\\
		\tan\beta &= \frac{w_\perp-w_\para'\tan\alpha}{u}.
	\end{split}
\end{equation}
From this then follows that
\begin{equation}
	\mc V_p\mc V_b\mc V\mc U H_1 \mc U^\dag\mc V^\dag\mc V_b^\dag\mc V_p^\dag = u\frac{\cos\alpha}{\cos\beta}b_1 - u\frac{\sin\alpha}{\sin\beta}p_1,
\end{equation}
to which we can apply our earlier methods.

The effects of these transformations on $H_0$ are then straightforwardly calculated. As usual, $\mc U(\vec u)$ gives us
\begin{equation}
	\mc U H_0 \mc U^\dag = \vec K\cdot\bs\gamma + m\gamma_4,
\end{equation}
which upon application of $\mc V$ turns into
\begin{equation}
	\begin{split}
		\mc V\mc U H_0 \mc U^\dag\mc V^\dag = K_1\gamma_1 + (K_2\cos\mu+K_3\sin\mu)\gamma_2&\\
		+ (K_3\cos\mu -K_2\sin\mu)\gamma_3 + m\gamma_4&.
	\end{split}
\end{equation}
With the identifications
\begin{align}
	q_1 &= K_1\cos\alpha + (K_2\cos\mu+K_3\sin\mu)\sin\alpha\\
	q_2 &= (K_2\cos\mu+K_3\sin\mu)\cos\alpha - K_1\sin\alpha\\
	q_3 &= (K_3\cos\mu -K_2\sin\mu)\cos\beta + m\sin\beta\\
	\tilde m &= m\cos\beta-(K_3\cos\mu -K_2\sin\mu)\sin\beta\\
	\tilde u &= u\frac{\cos\alpha}{\cos\beta}\\
	\tilde w &= - u\frac{\sin\alpha}{\sin\beta},
\end{align}
the Hamiltonian can be written as
\begin{equation}
	\mc V_p\mc V_b\mc V\mc UH\mc U^\dag\mc V^\dag\mc V_b^\dag\mc V_p^\dag = \vec q\cdot\bs\gamma + \tilde m\gamma_4 + \tilde ub_1 + \tilde wp_1, 
\end{equation}
from where we can refer back to the case with $\vec w\para\vec u$, which combined with our earlier discussion on parallel $\vec w$ and $\vec u$ gives us
\begin{equation}
	\begin{split}
		\mc F\mc V_p\mc V_b\mc V\mc UH\mc U^\dag\mc V^\dag\mc V_b^\dag\mc V_p^\dag\mc F^\dag = q_1\gamma_1 + \cos\delta(q_2\gamma_2+q_3\gamma_3)\\
		+ \tilde m\sqrt{1+\frac{\tilde w^2}{\tilde m^2}}\gamma_4 + \tilde ub_1 + \sin\delta(b_2q_3-b_3q_2),
	\end{split}
\end{equation}
on which we can apply the machinery developed for the case of $\vec w = 0$ (Eq.~\eqref{app:eq:hu}).

As a final comment on the general case, we note that the equations for the Weyl nodes now take the form
\begin{equation}\label{app:eq:kw}
	\bs\kappa(\vec k_W^{\pm}) = \pm r\hat{\vec u} + s(\hat{\vec w}\times\hat{\vec{u}}) + t(\hat{\vec u}\times(\hat{\vec w}\times\hat{\vec{u}})),
\end{equation}
where
\begin{align}
	r &= \frac{\cos^2\alpha}{\cos\beta}\sqrt{u^2\left(1-\frac{\tan^2\alpha}{\tan^2\beta}\right)-\frac{m^2}{\cos^2\alpha}}\\
	s &= \frac{m_1(\hat{\vec u}\times(\hat{\vec w}\times\hat{\vec{u}}))+m_2(\hat{\vec w}\times\hat{\vec{u}})}{w_\perp^2\sqrt{1-\hat{u}_3^2}}\cdot\hat{\vec z}\\
	t &= \frac{-m_1(\hat{\vec w}\times\hat{\vec{u}})+m_2(\hat{\vec u}\times(\hat{\vec w}\times\hat{\vec{u}}))}{w_\perp^2\sqrt{1-\hat{u}_3^2}}\cdot\hat{\vec z},
\end{align}
and,
\begin{align}
	m_1 &= m(\sin\alpha\cos\mu+\tan\beta\sin\mu)\\
	m_2 &= m(\sin\alpha\sin\mu-\tan\beta\cos\mu).
\end{align}
As two straightforward examples, we see that if $\vec w\para\vec u$, Eq.~\eqref{app:eq:kw} gives us
\begin{equation}
	\bs\kappa(\vec k^\pm_W) = \pm\sqrt{u^2-w^2-m^2}\hat{\vec u},
\end{equation}
while for $\vec w\perp \vec u$ it yields
\begin{equation}
	\begin{split}
		\bs\kappa(\vec k^\pm_W) = \pm\sqrt{1+\frac{w^2}{u^2}}\sqrt{u^2-m^2}\hat{\vec u}&\\
		+ \sgn[(\hat{\vec u}\times\hat{\vec w})\cdot\hat{\vec z}]\frac{mw}{u}(\hat{\vec w}\times\hat{\vec u}&).
	\end{split}
\end{equation}
Both of these cases reduce to the correct equation in the $\vec w = 0$ limit.

\section{ Realizations of Weyl metamaterials}\label{a:realizations}

To illustrate the utility of our theoretical framework and the physics of Weyl metamaterials, we consider three much-studied models as platforms for Weyl metamaterials. These include the topological insulator-ferromagnet layer structure [31], the popular toy model introduced by Vazifeh and Franz [29], and a Dirac semimetal model of Cd$_3$As$_2$ [9] with broken inversion symmetry. 
As detailed in Ref.~[31], the topological insulator-ferromagnet heterostructure can be modelled by the Hamiltonian
\begin{equation}\label{eq:heteroH}
	H = v_F\tau^z(\hat{\vec z}\times\bs\sigma)\cdot\vec k + \hat{\Delta}(k_z) + m\sigma^z,
\end{equation}
where $v_F$ is the Fermi velocity, $\vec k$ the momentum, $\hat{\Delta}(k_z) = \Delta_S \tau^x + \Delta_D(\cos k_z\tau^x-\sin k_z \tau^y)$, and $m$ is a time-reversal breaking magnetization. The parameters $\Delta_{S}$ and $\Delta_D$ characterize the tunnelling between different surfaces within and between neighboring layers of the heterostructure, respectively. While not unique, a suitable choice for the representation of the Dirac gamma matrices allows us to write Eq.~\eqref{eq:heteroH} as
\begin{align*}
	H = &v_F k_x\gamma_1 + v_F k_y\gamma_2 + \Delta_D\sin k_z d \gamma_3\\
	&+ (\Delta_S + \Delta_D\cos k_z d)\gamma_4 + \vec m\cdot\vec b,\TG
\end{align*}
where we have promoted the model from having magnetization only in the $z$ direction $mb_3\to \vec m\cdot\vec b$. Here we see that $\vec m$ corresponds to the TR-breaking field $\vec u$ in Table~1 of the main text and letting it vary as a function of position, we obtain an example of a TR-breaking metamaterial. By applying our general framework, it is straightforward to study the curved-space low-energy physics of the model. 

We next consider the toy model introduced in [29]. The model represents a 3d (topological) insulator with a finite magnetization and has a Hamiltonian
\begin{align*}
	H = &2\lambda\tau_z(\sigma_x\sin k_y - \sigma_y\sin k_x) + 2\lambda_z\tau_y\sin k_z + \tau_x M_{\vec k}\\
	&+  -u_0\sigma_y s_z + \vec u\cdot(-\tau_x \sigma_x, -\tau_x \sigma_y, \sigma_z)\TG,
\end{align*}
where we have slightly deviated from the notation in [29] for the TR and I breaking terms. Here, the mass term is $M_{\vec k} = \mu - 2t\sum_{\alpha = 1}^3 \cos k_\alpha$, where $\mu$ and $t$ are the chemical potential and hopping amplitude, respectively. The parameters $\lambda, \lambda_z$ characterize the strength of the spin-orbit hopping and $\vec{u}$ can be identified with magnetization. Expressing this Hamiltonian via Dirac gamma matrices gives us
\begin{align*}
	H = 2&\lambda(\gamma_2\sin k_y + \gamma_1\sin k_x) + 2\lambda_z\gamma_3\sin k_z\\
	+ &\gamma_4 M_{\vec k}+ u_0\varepsilon + \vec u\cdot\vec b. \TG
\end{align*}
Within the scope of our present formalism, we cannot account for the I-breaking $\varepsilon$ term and must hence restrict ourselves to $u_0 = 0$, but otherwise this is now in a form which we can apply our formalism to, starting from Eq.~\eqref{app:eq:hu}. For example, we immediately get that the equations for the Weyl nodes are given by
\begin{equation}
	2\lambda_j\sin k^W_j = \sqrt{u^2 - M_{\vec k^W}^2}\hat{u}_j,
\end{equation}
where $\lambda_{1,2} = \lambda$ and $\lambda_3 = \lambda_z$. The frame fields and the metric can be now extracted from our general formulas.  

Our third example is a 3D Dirac semimetal [9] with broken I symmetry. The Hamiltonian describes the spectrum of Cd$_3$As$_2$ (or Na$_3$Bi) near the $\Gamma$ point and is given by
\begin{equation}
	H_0 = \epsilon_0(\vec k) + M(\vec k) \tau_z\sigma_z + Ak_x\tau_z\sigma_x + Ak_y\tau_z\sigma_y,
\end{equation}
where $\epsilon_0(\vec k)=C_0+C_1k_z^2+C_2(k_x^2+k_y^2)$ and $M(\vec k)=M_0+M_1k_z^2+M_2(k_x^2+k_y^2)$ [9]. The above Hamiltonian can be directly translated into our gamma-matrix representation via the substitution $\tau_z\sigma_j = \gamma_j$ for $j = 1,2$ and $\tau_z\sigma_z = \gamma_4$. We can add an inversion-breaking field to this model by introducing a coupling $\vec w\cdot\vec p$, such that we get
\begin{equation}
	H = \varepsilon_0(\vec k) + Ak_x\gamma_1 + Ak_y\gamma_2 + M(\vec k)\gamma_4 + \vec w\cdot\vec p. 
\end{equation}
While the above form with the I-breaking vector parameter $\vec w$ is a bit abstract, it is possible to implement it by, for example, elastic deformations due to strain [22-25]. Starting from Eq.~\eqref{app:eq:hp}, we can then then derive the desired quantities, viz. the frame fields and the induced metric.

\section{Semiclassical motion and the geodesic equation}\label{a:semiclassical}

The geodesic equation can be derived from the semiclassical equations of motion in Eq.~(10) of the main text. In the absence of electromagnetic fields and Berry curvatures, the equations reduce to
\begin{align}
	&\dot{\vec{r}}=\partial_{\vec{k}}\mc E= \frac{\partial_{\vec k}(g^{ij}k_ik_j)}{2\mc E} \nonumber\\
	&\dot{\vec{k}}=-\partial_{\vec{r}}\mc E = -\frac{\partial_{\vec r}(g^{ij}k_ik_j)}{2\mc E},
\end{align}
which -- given that $\mc E$ is a constant of motion -- can by a rescaling of the time $t\to\varepsilon t'$ be brought into the form
\begin{align}
	&\dot{\vec{r}}= \partial_{\vec k}\frac{1}{2}g^{ij}k_ik_j \nonumber\\
	&\dot{\vec{k}}=-\partial_{\vec{r}}\mc E = -\partial_{\vec r}\frac{1}{2}g^{ij}k_ik_j.
\end{align}
We recognize these as the Hamilton equations of motion, which follow from extremizing the action
\begin{equation}
	S = \int_{t_i}^{t_f}\mc L(\vec r, \dot{\vec r}) dt
\end{equation}
with the Lagrangian given by
\begin{equation}
	\mc L = \dot{r}^nk_n - \frac{1}{2}g^{ij}k_ik_j = \frac{1}{2}g_{ij}\dot{r}^i\dot{r}^j.
\end{equation}
This problem is equivalent to extremizing the length of a path in a curved space given by the metric $g^{ij}$, and it immediately follows that $\vec r$ obeys the geodesic equation
\begin{equation}
	\ddot{r}^l + \tensor{\Gamma}{^l_i_j}\dot{r}^i\dot{r}^j = 0,
\end{equation}
with the connection $\Gamma$ given by
\begin{equation}
	\tensor{\Gamma}{^i_j_k} = \frac{1}{2}g^{il}(\partial_{r_j}g_{lk} + \partial_{r_k}g_{lj} - \partial_{r_l}g_{jk}).
\end{equation}

Reintroducing the Berry curvatures and the emergent magnetic field changes this picture: although the energy remains a constant of motion up to first order in spatial derivatives, the geodesic equation now becomes that of a charged particle in an external magnetic field. We have that
\begin{align}\label{app:eq:semiclass}
	&\dot{\vec{r}}=\partial_{\vec{k}}\mc E-\Omega_{\vec{k}\vec{ r}}\dot{\vec{r}}-\Omega_{\vec{k}\vec{ k}}\dot{\vec{k}}\nonumber\\
	&\dot{\vec{k}}=-\partial_{\vec{r}}\mc E+\Omega_{\vec{ r}\vec{k}}\dot{\vec{k}}+\Omega_{\vec{ r}\vec{r}}\dot{\vec{r}} - q\dot{\vec r}\times\tilde{\vec B}.
\end{align}
If we neglect terms that are second order or higher in spatial derivatives, we can remove $\Omega_{\vec r\vec r}$ and write for example $(\mathbb{I}-\Omega_{\vec r\vec k})^{-1} \approx \mathbb{I}+\Omega_{\vec r\vec k}$. The equations reduce to
\begin{align}
	\dot{r}^l &= (\delta^l_s-\Omega_{k_l r^s} + q\Omega_{k_l k^n}\varepsilon_{nsm}\tilde B^m)\partial_{k_s}\mc E +  \Omega_{ k_l k_s}\partial_{ r^s}\mc E\label{app:eq:rdot}\\
	\dot{ k}_l &= -\partial_{ r^l}\mc E - q\varepsilon_{lmn}(\partial_{k_m}\mc E)\tilde B^n.
\end{align}
By rewriting the first equation in component form and differentiating once more with respect to time, we have
\begin{align*}\label{app:eq:rdotdot}
	\ddot{r}^l &= \frac{d}{dt}\left[(\delta^l_s-\Omega_{k_l r^s} + q\Omega_{k_l k^n}\varepsilon_{nsm}\tilde B^m)\partial_{k_s}\mc E +  \Omega_{ k_l k_s}\partial_{ r^s}\mc E\right]\\
	&\approx \frac{1}{\mc E}\left[(\partial_{r^n}g^{il})k_i\dot{r}^n + g^{il}\dot{k}_i\right]\TG,
\end{align*}
which follows from using the fact that $\mc E = \sqrt{g_{ij}k_i k_j}$, and dropping terms with more than one spatial derivative. To show that this is equivalent to a geodesic equation to linear order, note that the left term on the right-hand side is already linear in spatial derivatives, and we can thus simply substitute $\vec k = \mc E \bs g^{-1}\dot{\vec r} + \mc O(\partial_{\vec r})$, which follows directly from Eq.~\eqref{app:eq:rdot} and once again dropping all higher-order spatial derivatives. Hence, Eq.~\eqref{app:eq:rdotdot} now becomes
\begin{align*}
	&\ddot{r}^l = \left[(\partial_{r^t}g^{lj})g_{js} + \frac{1}{2}g^{ln}(\partial_{r^n}g_{ts})\right]\dot{r}^s\dot{r}^t - \frac{q}{\mc E}g^{li}\varepsilon_{ijk}r^j\tilde B^k\\
	&\longrightarrow \ddot{r}^l+\tensor{\Gamma}{^l_s_t}\dot{r}^s\dot{r}^t = - \frac{q}{\mc E}g^{li}\varepsilon_{ijk}r^j\tilde B^k,\TG
\end{align*}
where
\begin{equation}
	\tensor{\Gamma}{^l_s_t} = \frac{g^{lj}}{2}\left(\partial_{r^s}g_{jt} + \partial_{r^t}g_{js} - \partial_{r^j}g_{st}\right)
\end{equation}
is the Christoffel symbol.

As a final remark, we point out that the above derivation used the fact that
\begin{align*}
	\frac{d}{dt}\mc E &= (\partial_{k_j} \mc E)\dot{k}_j + (\partial_{r^j} \mc E)\dot{r}^j\\
	&\approx (\partial_{k_j} \mc E)(-\partial_{ r^j}\mc E - q\varepsilon_{jmn}(\partial_{k_m}\mc E)\tilde B^n) + (\partial_{r^j} \mc E)(\partial_{k_j} \mc E)\\
	&= -q(\partial_{k_j} \mc E)(\partial_{k_m}\mc E)\varepsilon_{jmn}\tilde B^n = 0\TG,
\end{align*}
where the last equality follows from the antisymmetry of $\varepsilon_{jmn}$.

\section{Analytical solution for 3D lens geometry}\label{a:analytical}

In this appendix we present an analytical solution for the oscillatory behavior of the trajectories in the limit of small deviations from $x^2 + y^2 = 0$ and $k_x/k_z, k_y/k_z \ll 1$. The spatial part of the metric for the 3D lens is given by
\begin{widetext}
\begin{equation}\label{app:eq:lensm}
	\bs g =
	\begin{pmatrix}
		v_x^2(1-\frac{m^2}{u^2}\sin^2\theta(r)\cos^2\phi) & -\frac{1}{2}\frac{m^2}{u^2}v_xv_y\sin^2\theta(r)\sin2\phi & -\frac{1}{2}\frac{m^2}{u^2}v_xv_z\sin2\theta(r)\cos\phi\\
		 &  v_y^2(1-\frac{m^2}{u^2}\sin^2\theta(r)\sin^2\phi) & -\frac{1}{2}\frac{m^2}{u^2}v_yv_z\sin2\theta(r)\sin\phi\\
		 & & v_z^2(1-\frac{m^2}{u^2}\cos^2\theta(r))
	\end{pmatrix},
\end{equation}
\end{widetext}
where the omitted elements that can be deduced from $\bs g$ being symmetric. For small $\theta$, the metric reduces to
\begin{equation}
 	\bs g =
 	\begin{pmatrix}
 		v_x^2 & 0 & -\frac{m^2}{u^2}v_xv_z \omega x\\
 		0 & v_y^2 & -\frac{m^2}{u^2}v_yv_z \omega y\\
 		-\frac{m^2}{u^2}v_xv_z \omega x& -\frac{m^2}{u^2}v_yv_z \omega y& v_z^2(1-\frac{m^2}{u^2})
	\end{pmatrix}. 
\end{equation} 
From Eq.~\eqref{app:eq:lensm}, it is straightforward to calculate its spatial derivatives which in our approximation reduce to
\begin{align}
	\partial_x\bs g = -\frac{m^2}{u^2}
	\begin{pmatrix}
		2v_x^2\omega^2 x & v_xv_y\omega^2y & v_xv_z\omega\\
		v_xv_y\omega^2y & 0 & 0\\
		v_xv_z\omega & 0 & -2v_z^2\omega^2x
	\end{pmatrix}\\
	\partial_y\bs g = -\frac{m^2}{u^2}
	\begin{pmatrix}
		0 & v_xv_y\omega^2 x & 0\\
		v_xv_y\omega^2 x & 2v_y^2\omega^2 y & v_yv_z\omega\\
		0& v_yv_z\omega & -2v_z^2\omega^2y
	\end{pmatrix},
\end{align}
and $\partial_z\bs g = 0$. 

It follows from writing down the equations of motions using Eqs.~(10) in the main text using these approximations that the system of first order differential equations separate in $x$ and $y$, such that we have
\begin{widetext}
\begin{equation}
	\begin{split}
	\begin{pmatrix}
		\dot{x}^i\\
		\dot{k}_{x^i}
	\end{pmatrix}
	&=
	\frac{1}{\mc E}
	\begin{pmatrix}
		-\frac{m^2}{u^2}v_{x^i} v_z\omega k_z & v_{x^i}^2\\
		-v_zk_z\omega^2\left(v_zk_z\frac{m^2}{u^2} +cu(1-\frac{m^2}{u^2})^\frac{3}{2}\right) & \frac{m^2}{u^2}v_{x^i} v_z\omega k_z
	\end{pmatrix}
	\begin{pmatrix}
		x^i\\
		k_{x^i}
	\end{pmatrix},
	\end{split}
\end{equation}
\end{widetext}
for both of them. Here, we have ignored the contributions from the the Berry curvatures (which would couple the motion in the $x$ and $y$ directions) as their effects are small on short enough time scales. We have additionally taken into account the effective magnetic field arising from a position-dependent Weyl node, $\tilde{\vec B} = \nabla\times\vec k^W/q$. Since this magnetic field acts with different signs on the opposite nodes, we have introduced the parameter $c = \pm 1$ to distinguish the Weyl nodes ($c=0$ would correspond to pure geodesic motion). In the derivation of this linear differential equation, we have only kept terms up to linear order in $k_x$ and $k_y$ and used the fact that $\dot{k_z} = 0$. The solution to this differential equation can be formally written as $\Psi(t) = \exp(\bs A t)\Psi(t=0)$, where $\bs A$ is the matrix on the right-hand side. From this, it is evident that if $\bs A$ has imaginary eigenvalues, we will obtain oscillatory solutions. To simplify, we now assume that $v_x = v_y \equiv v$. The eigenvalues are then
\begin{align}
	\lambda^\pm = \pm i\sqrt{1-\frac{m^2}{u^2}}\sqrt{\frac{m^2}{u^2}v_z^2k_z^2 + c u\sqrt{1-\frac{m^2}{u^2}}}\frac{\omega v}{\mc E}
\end{align}
from which it follows that we have sinusoidal solutions typically only for one chirality $c = 1$. Furthermore, since $\dot{z} \approx v_z^2(1-m^2/u^2)k_z/\mc E$, we have that the length of one oscillation with period $T = 2\pi/|\lambda|$ is
\begin{equation}
	\Delta z = \dot{z}T  = 2\pi\frac{v_z}{v\omega}\sqrt{\frac{1-\frac{m^2}{u^2}}{\frac{m^2}{u^2}+\frac{u}{v_zk_z}\sqrt{1-\frac{m^2}{u^2}}}}.
\end{equation}
The distance between two focal points is $\Delta z/2$ which depends on the energy of particles weakly through $k_z$ in the square root. The analytical predictions are confirmed by exact numerics. The Berry-curvature effects that are neglected here affect the long time dynamics, most clearly by resulting in a non-planar motion that is not evident for short times. The analytical results are applicable also for different textures when $\omega$ is replaced by a characteristic measure of the variation of $u_z$ component near the lens axis. 

\begin{figure*}
\includegraphics[width=1.8\columnwidth]{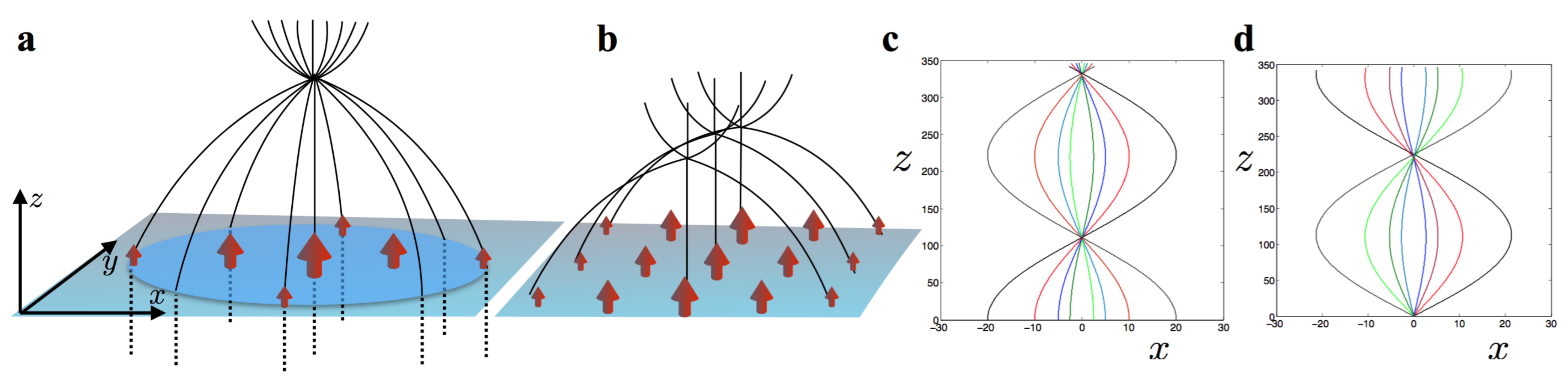} 
\caption{ \textbf{a}: 3d lens effect can be realized also by a unidirectional $\vec{u}$ texture where modulus increases away from the lens axis.  \textbf{b}: Unidirectional texture which varies only in $x$ direction realizes 2d lens effect in $x-z$ plane.  \textbf{c}: Semiclassical trajectories of carriers with texture like in \textbf{b} with $\vec{u}=\left[0,0, u(r)\right]$, $u(r)=u_0+u_1(1-e^{-x^2/\xi^2})$, $v_x=v_y=v_z=v$, $m/u_0=0.5$, $u_0/u_1=3$ and $\xi=100$. The numerical values of the coordinates are given in the units of $\xi$. The initial conditions for the trajectories are $y=z=0$, $k_x=k_y=0$, $k_z=0.6u/v$. Different curves correspond to different initial $x$ coordinates.  \textbf{d}: Trajectories of the same system but for initial conditions $x=y=z=0$, $k_z=0.6u_0/v$. Different curves correspond to $k_x=\pm0.016, \pm0.03,\pm0.07,\pm0.13u_0/v$.  } \label{supp}
\end{figure*}

\section{Alternative lens realizations}\label{a:lenses}
In the main text, we considered an electron lens in a two-node model $H=v_ik_i\gamma_i+ m \gamma_4+\vec u \cdot \vec b$ with a skyrmion-like magnetic texture. This is not necessary for the lens effect to occur as we can see in Fig. ~\ref{supp}, which illustrates how a similar effect can be realized by a unidirectional texture $\vec{u}=\left[0,0, u(r)\right]$.  If $u(r)$ decreases radially away from the lens axis as in Fig.~\ref{supp} \textbf{a}, the carrier trajectories near the axis are qualitatively similar to the rotating texture. However, even simpler textures result in nontrivial lensing; the case where $u(r)$ varies only in the $x$ direction, as shown in Fig.~\ref{supp} \textbf{b}, still leads to a 2d lens effect. It is encouraging from a practical point of view that no elaborate 3d textures are needed in realizing remarkable effects. In Figs.~\ref{supp} \textbf{c} and \textbf{d}, we have plotted the particle trajectories corresponding to a texture like in \textbf{b}. We have used a Gaussian envelope although the precise form is not crucial for the qualitative features. The motion in the $x-z$ plane looks qualitatively similar to the 3d-lens motion but only in that plane -- not for any arbitrary plane parallel to the $z$-axis as in a 3d lens.

\section{Numerical calculation of the local density of states}\label{a:ldos}

We employed the LDOS as a tool for comparing the exact four-band model and the linearized two-band Weyl block for inhomogeneous TR-breaking textures $\vec{u}(\vec{r})$.  This was done by considering a system that is inhomogeneous in the $z$-direction but homogeneous with periodic boundary conditions in the $x-y$ plane. The considered model has a Hamiltonian $H=\bs\kappa(\vec k)\cdot\bs\gamma + m \gamma_4+\vec u \cdot \vec b$, where $\kappa_i(\vec{k})=t\sin{k_i}$ and $m$ and $t$ are constants. In the numerics, we employed the representation $\gamma_i=\tau_z\sigma_i$ for $i=1,2,3$ and $\gamma_4=\tau_x$. The LDOS was studied for rotating and linear textures $\vec{u}(z)$. In the studied geometry, $k_x$ and $k_y$ are good quantum numbers so the LDOS can be expressed as $\nu(z,E)=\sum_{E_{k_x,k_y,n}}|\Psi_{k_x,k_y,n}(z)|^2\delta(E-E_{k_x,k_y,n})$, where $n$ denotes the discrete quantum number in $z$ direction. We considered a system on lattice with $N$ lattice sites in the $z$ direction, replaced $t\sin{k_z}$ with a standard hopping matrix, and imposed hard-wall boundary conditions. In the continuum model, the momentum operator $\partial_z$ was replaced by a hopping matrix representing a discretized derivative.  In order to evaluate the LDOS, we numerically calculated $E_{k_x,k_y,n}$ and $\Psi_{k_x,k_y,n}(z)$ for the 1d tight-binding models with a sufficiently dense $(k_x, k_y)$-lattice for the four-band model and the two-band models. To obtain continuous functions in energy, we replaced the energy delta functions by Lorentzians $\delta(E-E_n)\to \frac{1}{(E-E_n)^2+\eta^2}$ with broadening $\eta$.

\end{document}